# An agent-based model of modal choice with perception biases and habits

**Author(s) :**
Carole ADAM, Univ. Grenoble-Alpes, LIG, France – carole.adam@imag.fr
Benoit GAUDOU, Univ. Toulouse Capitole, IRIT, France - benoit.gaudou@ut-capitole.fr

## Abstract

**Introduction.**
To adapt cities to the issues of climate change and public health, urban policies are trying to encourage soft mobility [14] in order to reduce traffic and pollution, via financial incentives or new infrastructure. However, mobility evolves very slowly, and the share of the car remains significant (74% in France [9]), despite increased public awareness of global warming, and increased concern for ecology. The pandemic offered an opportunity to explore the impact of reduced car mobility and new urban planning policies, for instance with temporary cycle paths [19]. But these public policies normally take longer to implement and are not always well accepted by the car-loving population; many of these temporary cycle paths were gradually returned to cars after the end of the lockdowns [6]. Many explaining factors of this inertia of mobility and reluctance to shift from the car are already known, both contextual, such as a lack of alternatives (limited public transportation options), individual constraints (transporting children or tools), or higher costs of newer or electric vehicles...); and psychological, such as the difficulty to change habits [8, 17], individualism [12], or influence of cognitive biases [15, 13].
The SWITCH project[1] aims at developing prospective tools to reflect on scenarios for transition of cities towards more sustainable mobility [2, 1, 5]. In this context, we wish here to study the factors and obstacles, particularly psychological, of a modal shift for urban commutes. We developed an agent-based modal choice model integrating the influence of perceptual biases and habits, implemented in a Netlogo simulator. We conducted an online survey and collected 650 answers, which provided realistic values to initialise this simulator. We report here several scenarios run in this simulator to illustrate the impact of biases on the efficacy of public policies.

**Background.**
Modal choice has been studied in sociology under several axes: definition of user profiles, and analysis of modal choice criteria. The Mobil'Air study [10] presents 4 car user profiles with percentages of the population in each category but does not consider other modes of mobility. It also notes the importance of constraints, of the reason for travel (transporting children for example), and the strength of routines. Rocci [20] proposes a more detailed classification with 6 profiles including other modes than car, but without population distribution statistics. Similarly to Mobil'Air, her study also reports constrained or on the contrary passionate use of the car. Rocci also shows inter-individual differences in perception of modes: for example, convinced car drivers tend to underestimate its price and overestimate that of public transport. She defines the notion of 'mobility capital' as individual constraints for using a mode (owning a bicycle, having a driving license, being fit to cycle or walk, living close to a bus stop...). Beyond these constraints, everyone will evaluate different aspects, such as price, safety, or travel time. The choice must also minimise mental load (*e.g.* number of connections). Finally, we retained 6 decision factors [4]: cost, time, practicality, safety, comfort, and ecology.

Habits are essential in mobility decisions [8]: individuals tend to reproduce habitual decisions when in the same context. This process can save decision time, but also lead to misadapted decisions in an evolving environment. Habits can also modify perceptions: usual car drivers can overestimate travel time by public transport and under-estimate travel time by car [20, 7]. Life cycle changes that disrupt habits (job change, moving, birth, etc.) are favorable moments for breaking old habits and creating new ones [21]. Thus, the COVID-19 pandemic has shown an unusual reset of habits that encouraged bicycle travel, at least temporarily [6].

---

[1] Simulating the transition of transport Infrastructures Toward smart and sustainable Cities: https://www6.inrae.fr/switch

Cognitive biases are heuristics used by our cognitive system to facilitate decision-making [22], enabling fast reasoning during stressful or complex situations, despite incomplete or uncertain information. Although essential to our proper functioning, they can sometimes lead to irrational decision-making or errors. Innocenti *et al.* innocenti2013car show that people tend to 'stick' to the car, even when more expensive than public transport, and explain this irrationality by the influence of cognitive biases. They conclude that mobility policies must try to modify the perception of different modes of mobility. Another study [13] looks at the reasons why drivers are generally reluctant to switch from their personal car to new more efficient modes of mobility (carpooling or free-floating bikes and scooters). They find that mobility decisions are influenced by various emotions and cognitive biases not considered by mobility operators. The halo bias pushes motorists to amplify benefits of driving and ignore its disadvantages. The ambiguity bias pushes them to prefer known to unknown risks, thus avoiding uncontrollable risks posed by public transport. The anchoring bias pushes to retain a negative first impression and prevent future reuse of a new mode. The status quo bias induces a preference to keep things as they are to save cognitive load, similarly to habits.

**Agent-based model.**

Our approach is agent-based modelling: autonomous entities called agents represent individuals; the model describes their behaviour at the microscopic level; the behaviour of the society (macroscopic level) then emerges from the agents' interactions. Concretely, we will model individual modal choice decisions, and observe the resulting modal distribution in the global population. Our objective is to integrate psychological factors in these individual decisions, in order to illustrate their impact on the effectiveness of urban development policies. The model user will be able to modify the urban environment and observe how the modal distribution evolves in response, depending on whether the agent decisions are biased or not.

In our model, we consider 4 modes of mobility: car, bicycle, bus, or walking. They are evaluated according to the 6 criteria established above: ecology, comfort, price, time, practicality, and safety. The environment contains the objective values of the 4 modes with respect to the 6 criteria, representing the current urban infrastructure, and accessible to all agents.

The agents are endowed with individual priorities for the 6 criteria. They also have a filter which biases their perceptions; this filter contains a multiplying factor for each criterion of each mode (so 24 factors), modifying the objective value perceived in the environment. This filter is dynamic, computed as an average of the prototype filter for the usual mode and a neutral filter, weighed by the strength of habit: the more an agent is used to a mode, the more it biases its perception. Each agent therefore has its own subjective evaluation of the modes on all criteria (24 values), differing more or less from the objective values. The agent can then calculate the scores of the 4 modes following a multi-criteria evaluation formula [16]. Concretely, agent $i$ uses its priorities for each criterion $c$, denoted $prio(i,c)$, and its values of mode $m$ over each criterion $c$, denoted $val(m,c)$, to compute its score for mode $m$ as follows:

$$score(i,m) = \sum_{c \in crits} val(m,c) * prio(i,c).$$

The rational mobility choice for the agent is the mode receiving the maximal score. Each agent also has a home-work distance constraining its choices (walking available below 7km and cycling below 15km) and they may or may not have access to the bus and the car. In addition, the agent maintains a list of the modes used for its last journeys and uses it to deduce habits of each mode as their past frequency. This habit is then used as a probability to reuse the same mode without evaluation [1]; if no habit is triggered, modes are re-evaluated and the best one selected.

**Mobility survey.**

To realistically initialise the population of our model, we conducted a survey via an online form, sent to various university mailing lists (students, laboratories, research groups, etc.) or via our personal networks [11]. The questionnaire consists of three main parts. The first part concerns the responders profile (gender, home-work distance, number of weekly journeys, accessible modes) and their mobility habits (usual mode), without any identifying data. In a second part, participants are asked to report their priorities for the 6 decision criteria. The third part concerns their perceptions of the values of the mobility modes over these 6 criteria (24 scores). All ratings (priorities and values) are given on a Likert scale from 0 to 10. We collected 650 responses to this survey, between March and July 2023, available in open data [3].

The question about usual commuting mode allows us to deduce the following modal distribution in our sample: bicycle 31.38% (n=204), car 20.62% (n=134), bus 35.08% (n=228), walk 12.92% (n=84). In comparison, the national statistics for France [18] provide a very different distribution: bicycle 2%, car 74%, bus 16%, walk 6%. Our sample is therefore not representative at all, in particular because of the diffusion biases of the survey (academia), and geographical biases (diffusion from Grenoble, where the share of cycling is much greater than the national average[2]). However, the good coverage of each of the mobility modes studied will allow us to deduce interesting statistics.

The question about home-work distance provides distance statistics per usual mode. After removing aberrant responses (*e.g. w*alking 550km, driving 2000km, etc.) and zero distances, we obtain the following statistics: cyclists travel an average of 6.43 km (median 5km), pedestrians walk in average 1.8 km (median 1.5km), motorists drive in average 21.29km (median 15km), and bus trips are in average 11.16 km long (median 5.55km). Beyond the distance, not all modes are accessible to all users. We calculated the number of respondents using each mode and having expressed inaccessibility not due to distance. It is important to note that these are purely declarative and can be subjective. To simplify, we consider walking and cycling as accessible unless limited by distance. For the other 2 modes, we have found the following statistics: 57.46% of bus users, 29.9% of cyclists, and 50% of pedestrians cannot use a car; 61.19% of car drivers, 10.29% of cyclists, and 8.33% of pedestrians do not have access to public transport.

We have also computed the average priorities for the 6 criteria, in the total sample (n=650) and per usual mode. We observe marked differences between users of different modes, for example a very low priority of motorists for ecology (5.65 out of 10, compared to 7.08 in average) and price (5.63 out of 10, compared to 6.97 in average), of cyclists for safety (5.37 out of 10, vs 6.2 in average), or of pedestrians for time (6.7 out of 10, vs 7.47 in average). We then computed average evaluations of modes on criteria, in the total sample, and in users vs non-users of each mode. We noticed marked differences again, with users generally over-evaluating their mode compared to the rest of the population. Furthermore, these evaluation deviations are aligned with priority deviations: cycling is evaluated as very unsafe (4.62 in average, and even 4.28 over non-users), walking as very slow (2.98 in average, 2.69 over non-users), and driving as very expensive (2.68 in average, 2.38 over non-users, compared to 6.87 in average for the bus, which is the closest in terms of comfort) and not ecological (1.81 in average, 1.63 over non-users; compared to 9.81 in average for walking). From there, we calculated the average deviations between the value of each mode on each criterion, for its users, compared to the 'objective' value taken as the median over all responses (24 deviation factors); this constitutes the perception filter prototype specific to users of this mode.

**Modal choice simulator.**

We implemented a Netlogo [23] simulator of modal choices, accessible online[3]. The environment is extremely simplified, it contains neither buildings nor roads, but only the abstract numerical evaluation of the 4 modes over the 6 criteria, considered as the objective values. These are initialised from the survey: we calculated the median evaluation of all modes on all criteria, per usual mode; we then averaged these 4 different visions, weighed by the proportion of each mode in the French population [18] to adjust the average values in our non-representative sample. We created a population of 200 agents with this same distribution: 2% cyclists, 74% motorists, 16% bus users, and 6% pedestrians. Each agent is initialised with its usual mode, and corresponding attributes. Average priorities are computed over users of this mode, with an empirical random variation between -20 and +20% to obtain heterogeneous profiles. Its trip list initially contains only the usual mode, and its perception filter is the prototype for this mode. Its home-work distance is initialised by a random draw with a Gaussian distribution parameterised by the mean and standard deviation calculated for this mode, and determines its access to walking and bike. Its access to car and bus is drawn randomly according to the probability per mode. The number of agents created allows to smooth out the effect of randomness to observe macroscopic level effects.

---

[2] https://www.grenoble.fr/uploads/Externe/73/215_984_Grenoble-2e-ville-cyclable-de-France-pour-les-actif.pdf (press release) and https://www.insee.fr/fr/statistiques/2557426 (INSEE statistics)

[3] Online simulator: https://nausikaa.net/index.php/2024/05/28/mobility-switch-simulator/

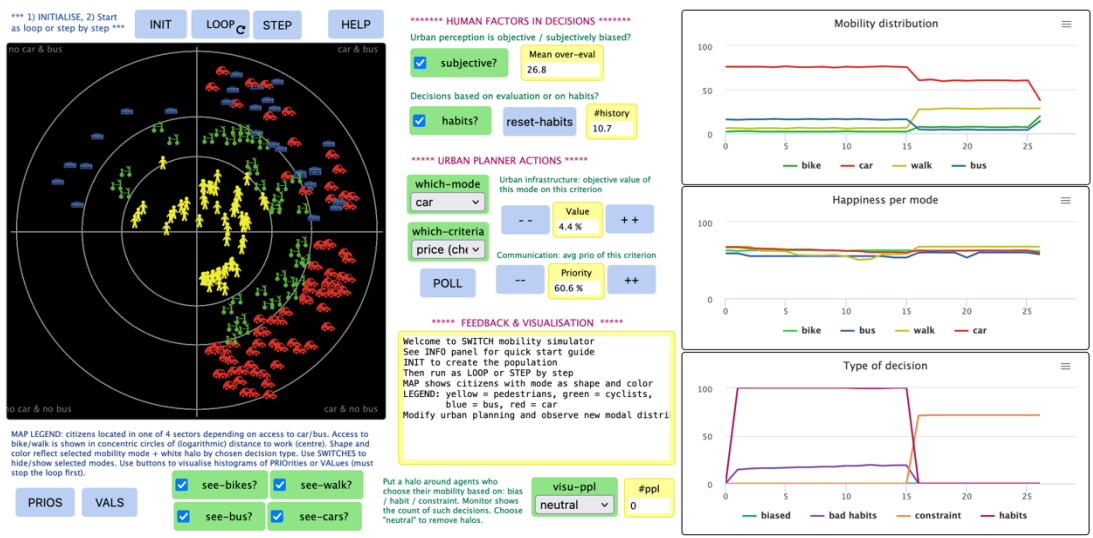

Figure 1: Netlogo simulator interface

Figure 1 shows the Netlogo simulator interface. It firstly allows the user to modify the urban layout, in an abstract way, by directly controlling the objective values of each mode on each criterion. These abstract modifications have the advantage of capturing all concrete urban policies (increasing cycling safety can correspond to building cycling lanes; decreasing car comfort can capture a reduction in parking spots), but have the disadvantage of also allowing modifications with no concrete meaning (one cannot really reduce the ecology of walking or make driving free). In the future, the simulator should provide the player with concrete actions choices, then translated into criteria values; but to explore the effect of biases, these abstract actions allow greater control. Along the same lines, the user can also modify the average priority of criteria (simulating communication campaigns, *e.g.* promote ecology or road safety); this is also a simplification to give more control and visualise the influence of these on macroscopic indicators. Finally, with the idea of exploring the impact of individuals' biases and habits, these two aspects of reasoning can be activated or deactivated by the user.

Once the population is initialised, each agent is represented with a shape and color corresponding to its current mobility. They do not move, we only model their decisions. At each time step, all agents perform a random draw to find out if they activate their habit or if they re-evaluate the 4 modes; in addition, random events (probability 1%) can prevent an agent from taking its usual mode, to simulate possible breaks in habits (car broken down, bus strike...). The user can also reset the habits of all the agents, to simulate a global crisis (*e.g. p*andemic): past journeys are deleted, and the frequencies reset to 0, forcing all agents to reevaluate the available modes rationally until they build up new habits. Urban layout (objective mode values) and priorities can be changed while the simulation runs. Several graphs allow the user to visualise the evolution of macroscopic indicators: modal distribution (percentage of agents using each mode), satisfaction (average score of each mode for its users); and decision counts (number of routine decisions, biased decisions, and constrained decisions).

**Experiments.**
In a first scenario, we want to illustrate the modal transfer when the environment evolves, and its obstacles, notably habit. We launch a simulation then gradually increase the safety of the bike. Initially the modal distribution remains stable, however the number of biased decisions increases (agents subjectively under-evaluating bicycle), as well as the number of constrained decisions (agents who would prefer bicycle but cannot access it due to distance). Gradually, some bus users convert to cycling but progress is very slow. If habits are reset during this evolution, some agents immediately shift from the 3 other modes to cycling; the reset forced them to re-evaluate modes and allowed them to realise whether cycling had become best. The number of routine decisions drops to 0, while the number of biased decisions increases (perception biases modify the evaluation) and then falls again (when new habits take over).

In a second scenario, we want to show the impact of constraints and habits on the modal shift. We launch the simulation without perceptual bias but with only habits to isolate their effect. We gradually reduce the comfort of the car to simulate the increase in difficulties (traffic jams, less parking spots, etc.). We observe a gradual shift of motorists with the shortest home-to-work distances towards walking. We notice that the number of forced choices increases, because motorists living further away are forced to continue using the car while its score decreases. If comfort is further reduced, motorists living further away then gradually switch to the bus, less well rated than walking but the only option. If habits are reset, the shift happens instantly. Ultimately, only constrained motorists remain, who continue to use the car because they cannot access the bus.

In a third scenario we want to show the impact of the perception filter. With the initial layout and filters enabled, the proportions of users of each mode remain stable, corresponding to the survey results. By disabling filters, we observe that the proportion of bus users decreases gradually (or even instantly if we also disable habits). Indeed, the initial layout is not very favorable to the bus (the plot shows bus users have the lowest satisfaction), often chosen by default because of distance or access constraints. The perception bias allows bus users to rationalise their choice *a posteriori* to improve satisfaction.

**Conclusion**

In this article, we presented a modal choice model, based on a multi-criteria evaluation, with individual priorities and biased evaluations, and integrating habits. We have described the results of a survey to calculate the parameter values of this model, these responses being published as open data [3]. We finally introduced a Netlogo modal choice simulator implementing this model and presented some use case scenarios. This simulator is extremely simplified but allows the visualisation of the impact of perception biases and habits, and therefore shows the importance of considering them in the development of urban development policies. Future work will enrich the simulator with other decision factors, such as social pressure.

**Acknowledgements**. The survey was carried out as part of Chloé Conrad's M1 internship [11]. This work is funded by the ANR (French National Research Agency) as part of the SwITCh project, under number ANR-19-CE22-0003.